# Tunable spin polarization in III-V quantum wells with a ferromagnetic barrier

R. C. Myers, A. C. Gossard, and D. D. Awschalom

*Center for Spintronics and Quantum Computation, University of California, Santa Barbara, CA 93106*

Abstract

We demonstrate the epitaxial growth of optical-quality electrically-gated III-V ferromagnetic quantum structures. Photoluminescence spectroscopy reveals that initially unpolarized photoexcited holes in a GaAs quantum well become spin-polarized opposite to the magnetization of an adjacent digital ferromagnetic layer in the $Al_{0.4}Ga_{0.6}As$ barrier. A vertical bias is used to tune the spin polarization from –0.4% to 6.3% at T = 5 K and B = 1 kG during which the luminescence becomes quenched, indicating that the polarization is mediated by wave function overlap between heavy holes in the quantum well and Mn-ions in the barrier. Polarization is observed under negligible current flow and is insensitive to the initial spin orientation of the carriers, differentiating the effect from both electrical and optical spin injection.

Much recent work has focused on studying the interactions between free carrier spin and magnetic ion spin in semiconductors. An accurate knowledge of the strength, spatial extent, and sign of these interactions in magnetic semiconductors would offer a significant advantage in the field of spintronics for predicting the properties of carrier mediated ferromagnets. It would be desirable to engineer a heterostructure in which the strength of the interaction between free carrier spin and magnetic ion spin can be studied locally; such structures have been studied in II-VI systems in which paramagnetic digital layers of Mn were deposited within a quantum well.[1] Spin-LEDs have recently become the structures of choice for studying spin injection processes in magnetic semiconductors,[2] but due to spin-scattering during transport the polarization measured in these devices does not directly reflect the local interaction between the spins of the ferromagnetic ions and those of the free carriers. Therefore, in an effort to develop an architecture where the local interaction between magnetic ion spin and free carrier spin can be studied, we have engineered heterostructures in which the direct spatial overlap between magnetic ion spin and free carrier spin can be controlled electrically and probed optically.

Here we describe the optical characterization of III-V quantum wells (QW) with a ferromagnetic barrier grown by a combination of high temperature (HT) molecular beam epitaxy (MBE) and low temperature (LT) atomic layer epitaxy (ALE).[3] Free carrier spin polarization is observed in which initially unpolarized photoexcited holes in a GaAs QW become spin-polarized through interaction with an adjacent digital ferromagnetic layer in the $Al_{0.4}Ga_{0.6}As$ barrier. This spin polarization is measured through the polarization of the photoluminescence (PL), which is seen to qualitatively track the magnetization of the ferromagnetic layer as a function of both field and temperature. We apply a vertical bias in the regime of negligible current flow, resulting in a distortion of the potential and hence displacement of the wave function in the absence of charge injection. The fact that under these conditions we observe a modulation of the PL polarization suggests that the polarization arises from exchange interaction between heavy hole (HH) spins in the ground state of the QW and the spins of the Mn ions in the barrier. This conclusion is further supported by the antiparallel alignment of the hole and Mn spins observed under negative bias.

A schematic of the sample structures is shown in Fig. 1(a). The samples are grown using two Varian/EPI GEN-II MBE chambers. In chamber A, optimized for high mobility III-V's, the gated QW structure is grown by MBE at high temperature[4] and consists of the following layers: 300 nm GaAs buffer / 180 nm superlattice 30×(3 nm AlAs / 3 nm GaAs) / 200 nm n-GaAs (Si: n = $1\times10^{18}$ cm$^{-3}$) / 200 nm GaAs / 500 nm LT $Al_{0.4}Ga_{0.6}As$ / 350 nm $Al_{0.4}Ga_{0.6}As$ / 7.5 nm GaAs QW / $d$ $Al_{0.4}Ga_{0.6}As$ (where $d$ is either 5 nm or 9 nm).[5] The samples are subsequently cooled to room temperature and capped with As for in-air transfer to chamber B for LT magnetic overgrowth. The magnetic layers consist of digital ferromagnetic heterostructures (DFH)[6] grown by ALE, in which the composition of each monolayer is controlled by sequentially depositing each constituent element with sub-monolayer precision allowing for digital alloying within a single monolayer (ML) and the formation of thin ferromagnetic layers of MnAs.[7] Three different magnetic structures are grown where in each structure the first MnAs layer is deposited at a spacing $d$ from the edge of the QW and the magnetic layer is capped by 177 ML $Al_{0.4}Ga_{0.6}As$ / 27 ML GaAs also grown by ALE. Samples with magnetic structure 5×(0.5/20) and 5×(0.3/20) consist of 5 periods of DFH superlattice of Mn-rich layers (0.5 ML and 0.3 ML of MnAs, respectively) spaced by 20 ML $Al_{0.4}Ga_{0.6}As$. Samples with magnetic structure 1×(0.5/84) consist of a single Mn-rich layer, 0.5 ML MnAs / 84 ML $Al_{0.4}Ga_{0.6}As$. Finally, a Mn-free control sample consisting of 85 ML of ALE grown $Al_{0.4}Ga_{0.6}As$ is prepared. For gating measurements, we use standard photolithography and wet etching followed by In soldering to obtain Ohmic contact to the bottom layer of n-doped GaAs to serve as the back gate. A transparent front gate, which consists of 50 Å Ti / 40 Å Au is evaporated on the sample surface. With the 500 nm LT $Al_{0.4}Ga_{0.6}As$ serving as an insulating layer, we apply a voltage bias ($V_b$) defined as front minus back voltage. The structures can be biased from –15 V to +2 V with current flow less than 10 μA corresponding to a current density of less than 0.25 mA/cm$^2$. Table I summarizes the structure and preparation of the samples.

Photoluminescence intensity and polarization are measured in the Faraday geometry (magnetic field parallel to optical pump and collection path) using linearly polarized light from a mode-locked Ti/Saphire laser with an energy of 1.731 eV and intensity of ~56 W/cm$^2$. We define polarization as: $P = (RCP - LCP) / (RCP + LCP)$, where RCP and LCP are the intensities of right circularly polarized and left circularly polarized luminescence, respectively.

The magnetic properties of the samples are measured using a superconducting quantum interference device (SQUID) magnetometer and are shown for a representative sample (A), Fig. 1(b) and (c) (black data). As compared to DFH grown with GaAs spacer layers,[3,6] the samples with $Al_{0.4}Ga_{0.6}As$ spacer layers show markedly different magnetic behavior. In particular, magnetic hysteresis appears with field applied out-of-plane for $Al_{0.4}Ga_{0.6}As$ DFH, while in GaAs DFH this direction is a magnetic hard axis showing no hysteresis. Square hysteresis is also not observed in-plane, indicating that the anisotropic easy axis may lie along a non-trivial crystal direction. The Curie temperature ($T_c$) of these structures is ~15 K compared with ~40 K for GaAs DFH. Similar rotation of anisotropy and decrease in $T_c$ for the case of random alloy (Al,Ga,Mn)As are reported.[8]

Here we discuss the PL intensity and polarization for sample A at a fixed $V_b$ (-3.8 V). Figure 1(d) plots the PL intensity and polarization spectra of the QW at three different magnetic fields, revealing large field dependent polarization unaccompanied by a spectral shift in the PL. Magnetic field and temperature dependence are extracted by integrating the polarization over the full range of the PL spectrum [indicated by dashed lines in Fig. 1(d)] at each value of applied field and temperature, Fig. 1(b) and (c) (red circles), respectively. The polarization of the PL tracks the magnetization of the adjacent ferromagnetic layer indicating that the photoexcited carriers in the QW become spin-polarized through interaction with the magnetic layer before recombining. The decrease in hysteresis in the polarization data is to be noted, but could be due to the effects of illumination on the magnetism of the ferromagnetic layer[9] or discrepancies between the magnetic response of the as grown sample (measured in the SQUID) and processed device (measured in PL). The PL polarization of the control sample (red line) shows a weak linear field dependence consistent with the Zeeman effect at low magnetic field. To test the spatial extent of the polarizing interaction, we investigate sample B in which four additional 0.5 ML MnAs layers are inserted at 20 ML (5.7 nm) spacings. We observe no qualitative difference in the field dependence of the PL polarization between samples A and B, while minor variations in absolute polarization intensity are within the observed noise of sample reproducibility.[10] This variation could arise from, for example, sensitivity to the quality of the interface between the MBE and ALE grown regions of the sample. The fact that the additional MnAs layers in sample B produce no noticeable difference in the polarization behavior indicates that only the ferromagnetic layer closest to the QW is active in the polarization phenomenon. Moreover, the spacing of the second nearest magnetic layer to the QW, which for sample B is an effective $d$ of 10.7 nm, serves as an upper limit on the spatial extent of the polarizing interaction. We therefore conclude that the observed interaction occurs within 10.7 nm of the QW, consistent with the lack of polarization seen in unbiased samples with large $d$ values (samples D and E).[10] Further, we can also rule out spurious path-dependent optical polarization effects such as magnetic circular dichroism (MCD).[11] In the case of MCD polarization would occur via luminescence scattering in the magnetic layers as the light travels from the QW to the sample surface, thus MCD should scale with the total thickness of magnetic material, which is not the case here.

Both quantum confinement and strain significantly alter the selection rules in a QW, such that HH spins are pinned along the growth direction, while light hole (LH) spins are pinned in-plane[12] resulting in PL polarization that depends on collection geometry. Accordingly, we measure QW luminescence in the edge-emission geometry (field in-plane) and observe no polarization,[10] which suggests that the PL polarization arises from hole spin polarization since electron spin polarization should be isotropic (s-like). In contrast, hole spin polarization ought to exhibit large anisotropy such that in the edge-emission only LH spin polarization would result in an observed PL polarization, whereas in the surface emission (field out-of-plane) only HH spin polarization would result in an observed PL polarization. These results suggest that the PL polarization results from recombination between unpolarized ground state electrons in the QW conduction band and spin-polarized holes in the ground state of the valence band (HH).

Figure 2 shows the bias and spectral dependence of the PL intensity and polarization for the control sample (a) and for sample A (b) at a fixed magnetic field of +1 kG. In the control sample we observe the quantum confined stark effect (QCSE) such that at negative bias the PL is shifted to lower energy,[13] while no significant polarization is observed (<0.5%) at any bias. The same QCSE red shift is observed in the magnetic sample (A), however the intensity of the PL decreases at high negative bias coinciding with a region of large (2-8%) polarization. Qualitatively identical behavior is observed in the other magnetic samples under bias. Comparison with the control sample indicates that the PL quenching is a result of non-radiative recombination caused by interaction with the Mn layer in the barrier and not due to generic LT growth related defects. Because the laser pump (1.73 eV) is tuned below the band gap of the $Al_{0.4}Ga_{0.6}As$ barriers (~1.92 eV), the photoexcited carriers are confined to the QW. This fact together with the lack of significant current flow imply that the PL quenching is caused by carrier tunneling through the barrier resulting in non-radiative recombination with defect states in the magnetic layer. This indicates that the HH wave function is overlapping considerably with the magnetic layer.

Solutions to the one-dimensional Poisson equation for the valence band edge along the growth axis of Sample A are shown in the bottom panel of Fig. 3(a) at several biases; the corresponding ground state HH wave functions are shown in the top panel.[14] At $V_b = 0$ V, large band bending due to acceptors results in a triangular distortion of the square potential shifting the center of the wave function toward the magnetic layer. By applying a negative bias, the HH wave function overlap with the Mn ions is further increased, which in turn leads to tunneling, quenched PL, and hole spin polarization as indicated by the preceding analysis of Fig. 2. The bias dependence of the PL polarization for the gated sample set is plotted in Fig. 3(b). For all three magnetic samples there is a cross-over bias voltage ($V_{cb}$) at which the polarization changes sign. Below $V_{cb}$, the polarization increases to its maximum value while a large QCSE red shift of the PL is observed. Above $V_{cb}$, the polarization decreases below zero however no QCSE shift occurs. The fact that $V_{cb}$ cannot be defined for the control sample (its polarization shows no sensitivity to bias) allows us to attribute the cross-over phenomenon to an effective coupling between HH spin in the QW and Mn-ion spin in the barrier.

For a representative magnetic sample (A) the magnetic field dependence of the PL polarization is measured for numerous biases and illustrative results are shown in Fig. 3(c). For $V_b \leq -1$ V the polarization shows a positive field dependence, while for $-1$ V $< V_b$ the polarization shows a negative field dependence, thus $V_{cb} \sim -1$ V for this sample. In our measurement geometry, a positive polarization at a positive magnetic field corresponds to angular momentum of the emitted photons pointing anti-parallel to the magnetic field. The Mn-ion spins will align parallel to the magnetic field; this indicates that the net angular momentum of the recombining HH excitons is oppositely aligned to the Mn-ion spins in the bias range of $V_b \leq V_{cb}$, whereas above this bias the opposite is true. Additionally, the HH exciton spin is parallel to the HH spin. We therefore determine that the effect of $V_b$ is to flip the sign of the effective coupling between HH spin and Mn-ion spin from antiparallel in the case of $V_b \leq V_{cb}$, to parallel for $V_b > -V_{cb}$. From Fig. 3(b), it is clear that the bias dependence can vary between samples grown on the same day under nominally identical growth conditions. However, in all samples studied, the largest polarization is observed with $V_b < V_{cb}$ exhibiting an anti-parallel effective coupling between hole and Mn-ion spins, the sign of which is consistent with the antiferromagnetic coupling expected between free-holes and Mn-ion spins in III-V materials.[15]

We also study the PL intensity and polarization for sample A while optically injecting spin using a circularly polarized pump beam. The bias dependence of the PL intensity shows no sensitivity to the polarization state of the pump beam.[10] For a RCP versus LCP pump beam the polarization is increased and decreased respectively showing that optical spin injection has been achieved [Fig. 3(d)]. By simple averaging of the polarization under RCP and LCP illumination, the voltage dependence of the polarization matches the case of the linearly polarized pump beam. This indicates that the spin polarization mechanism shows no sensitivity to the initial spin state of the interacting carriers, such that optical spin injection can be seen as a simple shift of the overall PL polarization magnitude without changing the strength of the HH and Mn-ion spin interaction.

Finally, room temperature Hall measurements are carried out with samples prepared in the Van der Pauw geometry using soldered In for electrical contact. Values of two-dimensional carrier concentration and mobility are presented in Table I. Ohmic contact is achieved in several samples indicating modulation hole doping from the adjacent Mn-rich layer into the QW forming a two-dimensional hole gas. For magnetic samples in which Ohmic contact is not achieved, a non-linear I-V indicative of hopping conductivity through the DFH layers is observed.[3] No correlation between hole concentration and spin polarization is observed.

In summary, we have achieved large and electrically-gated hole spin polarization at low magnetic field in optical-quality III-V ferromagnetic quantum structures without the use of optical or electrical spin injection. We conclude that the spin polarization mechanism is highly local (<11 nm) being mediated by wave function overlap between HH in the QW and Mn-ions in the barrier. By shifting the HH wave function using a vertical bias, we are able to qualitatively vary the strength of the polarizing interaction, while at a certain bias the effective coupling between hole and Mn-ion spin changes from anti-parallel to parallel.

We thank E. Johnston-Halperin for help in the design of the original structure and for suggestions, J. English for MBE technical assistance, R. Seshadri for additional SQUID measurements, Y. Kato, M. Poggio, and A. Holleitner for helpful discussions. This work was supported by DARPA and the ONR.

desorption. The substrate was then cooled to 230°C, monitored by band edge thermometry, for ALE overgrowth. The growth rates for Ga, Al, and Mn were ~0.3 ML/s, ~0.17 ML/s, and ~0.02 ML/s respectively, as calibrated by RHEED oscillations. The arsenic growth rate was calibrated as specified in reference 3.

Figure captions:

TABLE I. Structure and preparation of all samples discussed. Column "P" indicates if spin-polarization was observed. Data from room temperature Hall measurements is provided indicating whether or not Ohmic contact was achieved (linear I-V) and the two-dimensional hole concentration ($p_{2D}$).

FIG. 1. (a) Schematic of sample structure (not to scale), electrical wiring, and measurement geometry. Cone indicates the direction of PL surface emission. Arrow "PL pump" shows the path of the pump beam and the direction of the applied magnetic field (B). (b) Magnetization of sample A (without bias) and PL polarization (P) at $V_b = -3.8$ V as a function of B. Open and closed symbols indicate the direction of field sweep as up and down, respectively. Control sample at the bias value of maximum polarization is included for comparison (red line). (c) Magnetization of sample A (without bias) and PL polarization at $V_b = -3.8$ V as a function of temperature (T). (d) Polarization and PL intensity spectra for QW at three field values. Dashed lines indicate the bounds of integration used to calculate the values of polarization presented in (b), (c) and Fig. 3.

FIG. 2. Spectral dependence of voltage tunable spin-polarization at 5 K and +1 kG for (a) non-magnetic control sample and (b) sample A. Top panels show the PL intensity, while corresponding PL polarization is plotted in the bottom panels.

FIG. 3. (a) Valence band edge diagram at three values of $V_b$ (bottom panel) and corresponding $HH_1$ wavefunctions (top panel). (b) PL polarization as a function of $V_b$ for biased samples. (c) Field dependence of sample A at various values of $V_b$. Open and closed symbols indicate the direction of field sweep as up and down, respectively. Control sample at the bias value of maximum polarization is included for comparison (blue line). (d) PL polarization as a function of $V_b$ for sample A under optical excitation with different helicities. The data for RCP and LCP excitation are averaged (RCP+LCP)/2 and compared with the case of zero optical spin-injection (linear pump).

| Sample | d (nm) | magnetic layer | gated | P | Ohmic | $p_{2D}$ (cm$^{-2}$) |
|---|---|---|---|---|---|---|
| A$^\$$ | 5 | 1×(0.5/84) | yes | yes | no | - |
| B$^\$$ | 5 | 5×(0.5/20) | yes | yes | no | - |
| C$^\$$ | 5 | 5×(0.3/20) | yes | yes | no | - |
| control$^\$$ | 5 | - | yes | no | no | - |
| D$^\#$ | 9 | 1×(0.5/84) | no | no | yes | $8.58\times10^{11}$ |
| E$^\#$ | 9 | 5×(0.5/20) | no | no | yes | $2.08\times10^{12}$ |
| F$^*$ | 5 | 1×(0.5/84) | no | yes | yes | $9.67\times10^{11}$ |
| G$^*$ | 5 | 5×(0.5/20) | no | yes | yes | $1.56\times10^{12}$ |

$^{\$,\#,*}$ Indicate samples from same template and growth day.

Table I
Myers *et al.*

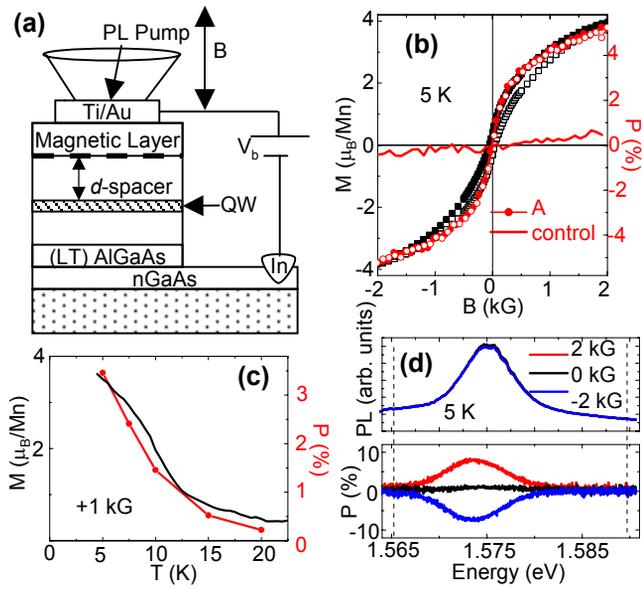

Figure 1
Myers *et al.*

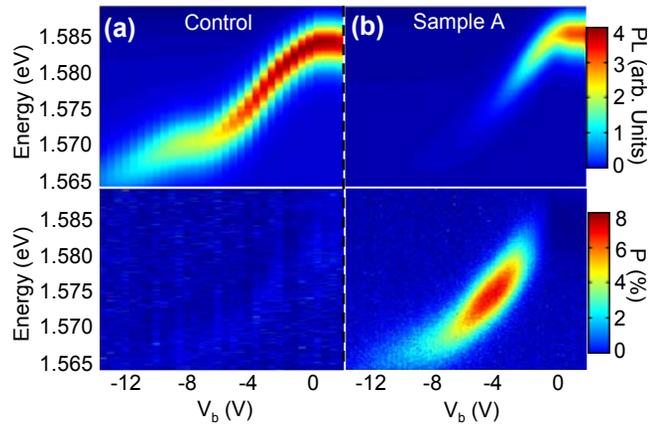

Figure 2
Myers *et al.*

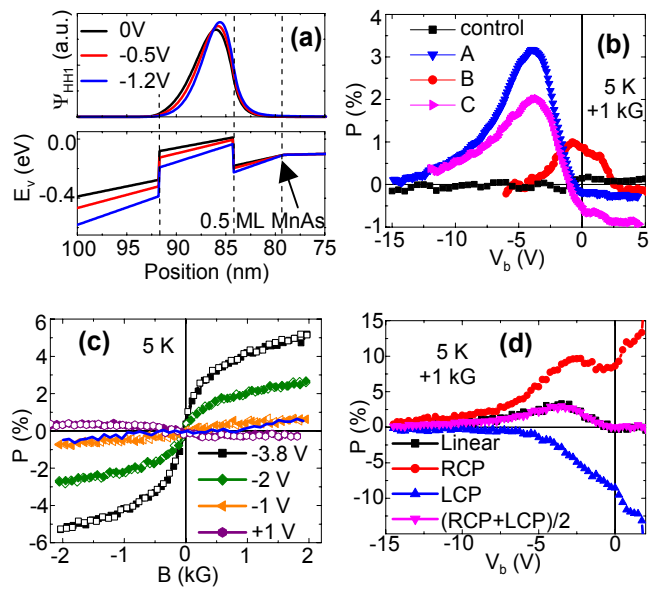

Figure 3
Myers *et al.*